\documentstyle[preprint,aps]{revtex}
\begin{document}
\draft

\title{Updated resonance photo-decay amplitudes to 2~GeV  }
\author{Richard~A.~Arndt, Igor~I.~Strakovsky$^\dagger$ and Ron~L.~Workman}
\address{Department of Physics, Virginia Polytechnic Institute and State
University, Blacksburg, VA~24061}

\date{\today}
\maketitle

\begin{abstract}

We present the results of an
energy-dependent and set of single-energy partial-wave analyses of
single-pion photoproduction data.  These analyses extend from threshold
to 2~GeV in the laboratory photon energy, and update our previous analyses
to 1.8~GeV.  Photo-decay amplitudes
are extracted for the baryon resonances within this energy range.
We consider two photoproduction sum rules and
the contributions of two additional resonance candidates
found in our most recent analysis of $\pi N$ elastic scattering data.
Comparisons are made with previous analyses.

\end{abstract}


\narrowtext
\section{INTRODUCTION}
\label{sec:intro}

The study of baryon resonances and their electromagnetic decays is
expected to see a resurgence as CEBAF becomes fully operational.
The resulting flood of very precise photo- and electro-production data
will provide a challenge for data analysts and model-builders alike.
For many of these reactions, single-pion photoproduction serves as
a "bench mark". In addition to providing the photo-decay amplitudes for
resonances with appreciable couplings to the $\pi N$ channel, it
fixes the $Q^2 = 0$ point for  pion electroproduction analyses. The
amplitudes from these analyses have also been utilized in evaluating a
number of sum rules which test the predictions of Chiral Perturbation
Theory (ChPT) and extended current algebra.
We will briefly discuss the status of two such sum rules
in  Section~IV.

The present analysis is a significant improvement on our previously
published result\cite{larw}
for three main reasons. The database was carefully
reexamined in order to check the assignments of statistical and
systematic errors. This resulted in a number of changes which are
discussed in Section II. The upper limit of our energy range was
increased from 1.8~GeV to 2~GeV (in order to better regulate the
solution near 1.8~GeV). Finally, the effect of two new resonance
candidates was considered. These were found in a search for possible
"missing resonances", as described in our most recent analysis\cite{aswp}
of elastic $\pi N$ scattering data.
In Section~III, we give the results of our multipole analyses as well as
the photo-decay amplitudes for resonances within our energy
region. Finally, in Section~V, we summarize our results and consider what
improvements can be expected in the future.

\section{The Database}
\label{sec:dat}

The data compilations of Ukai and
Nakamura\cite{uk85} and an earlier compilation by Menze, Pfeil and
Wilcke\cite{me77} were the main sources used in constructing our database.
In the present study we have attempted to verify all references contained in
these earlier compilations (more than 200 papers).
As a result, we have corrected some data, photon energies,
and systematic uncertainties according to the publications and/or
the authors' suggestions. For example, the threshold $\pi ^{0}p$
differential cross sections produced by MAMI at Mainz now consist of 11
angular points (instead of 21), and have an increased
normalization factor of 7.5\% \cite{be90}. Total cross sections produced
by ALS at Saclay were corrected by a factor of 0.94\cite{mz86}.
We have added some data missed in our previous
analyses\cite{larw},\cite{ar901}. Other data were removed
when found to be duplicated in our database
or according to the authors' suggestions. A small number of points were
removed when no reliable source was found.

As in our previous analyses, not all of the
available data were used. Data taken before 1960 were not analyzed,
nor were those single-angle and single-energy points measured prior
to 1970.  Some individual data points were also removed
from the analysis in order to resolve database conflicts.
Our previous published pion photoproduction scattering analysis\cite{larw}
(SP93) was based on 4015 $\pi ^{0}p$, 6019 $\pi ^{+}n$, 2312 $\pi ^{-}p$,
and 120 $\pi ^{0}n$ data.  Since then we have added 698 $\pi ^{0}p$ and 351
$\pi ^{+}n$ data. Through the checks described above, the total number of
$\pi ^{-}p$ data actually decreased by 96.

The new low-energy data have been produced mainly by TRIUMF,
for radiative pion capture on protons (39 differential and 10 total
cross sections\cite{wa92} and 10 measurements of P\cite{st94}),
and by LEGS at BNL for $\pi ^{0}p$  photoproduction
(12 differential cross sections and 97 measurements of
$\Sigma$)\cite{sa94p}. We also have 9 $\Sigma$ measurements
from LEGS for
$\pi ^{-}p$ photoproduction\cite{sa95p}, and a small number of SAL
$\pi ^{+}n$  photoproduction data (16 differential and
3 total cross sections)\cite{fi94p}, \cite{br95}.

Medium-energy
differential cross section data for $\pi ^{+}n$ (245 data)\cite{bu94} and
T measurements for $\pi ^{+}n$ (216 data) and $\pi ^{0}p$
(52 data) were produced by ELSA at Bonn\cite{an94p}.  We have
added 18 missing polarization measurements for $\pi ^{0}p$ in the
1~GeV region from Yerevan\cite{av91} and
$\pi ^{0}p$ (7 P\cite{bv86} and 14 $\Sigma$ \cite{gb771}, \cite{gb772}),
$\pi ^{+}n$ (18 $\Sigma$, P, and T)\cite{ge89},
and $\pi ^{-}p$ (16 T)
\cite{ag891} measurements between 230 and 700~MeV from Kharkov.
The distribution of
recent (post-1993) data is given in Fig.~1.

Other experimental efforts will soon provide data in the low to intermediate
energy region.  These include a
precise measurement of $\pi ^{0}p$ differential cross
sections made in a LEGS experiment.
This experiment spanned the $\Delta$
isobar region and was completed at BNL in 1992\cite{sa951}.  The
region between 145 and 200~MeV was covered
by MAMI at Mainz in 1991\cite{be94p}, and the
first phase of a measurement from threshold to about 25~MeV
has been completed at SAL\cite{hu95p},\cite{be95p}.
A double polarization (beam--target) experiment at PHOENICS
below 1150~MeV is planned at Bonn\cite{an941}.  We also expect that
the 1 to 2~GeV region
will be extensively studied at CEBAF\cite{cebaf1},\cite{cebaf2}.

\section{MULTIPOLE ANALYSES AND PHOTO-DECAY AMPLITUDES}
\label{sec:mul}

As in our previous studies\cite{larw},\cite{ar901}, we have performed
both energy-dependent and single-energy analyses. The single-energy
analyses were done mainly in order to check for structure missing in the
energy-dependent form. However, these results were also used in Breit-Wigner
fits to extract photo-decay amplitudes, as described below.
The methods used to generate these solutions
have been discussed previously\cite{larw},\cite{ar901}.
In the present analysis, one further degree of freedom was allowed.
Some multipoles were given an overall phase $e^{i\Phi}$ where the
angle $\Phi$ was proportional to $({\rm Im} T_{\pi N} - T_{\pi N}^2 )$.
This form satisfies Watson's theorem for elastic $\pi N$
amplitudes ($T_{\pi N}$) while exploiting the undetermined phase for
inelastic amplitudes.

Our energy-dependent solution (SM95) has a $\chi^2$ of 31810 for 13415
data to 2 GeV. The overall $\chi^2$/datum (about 2.4) is considerably
lower than that found in our previously published\cite{larw}
analysis to 1.8 GeV. While the
number of data has increased by about 1000 points, the $\chi^2$/datum has
decreased significantly from the value (3.6) reported for the SP93
energy-dependent solution. This result mainly reflects the database changes
discussed in Section~II. Our present and previous solutions are compared
in Table~I.

The very low energy region is complicated by different thresholds for
$\pi^0 p$ and $\pi^+ n$ production. While we have obtained a reasonable
fit to the available differential and total cross sections, the multipole
amplitudes should not used in the $\pi^+ n$ threshold region.
We have concentrated on the
extraction of resonance parameters, whereas the threshold region requires
a detailed study.

The results from our first analysis\cite{larw} (SP93) to 1.8 GeV are
compared with the present (SM95) energy-dependent and single-energy
multipoles in Fig.~2. Significant deviations from SP93 are visible
in multipoles connected to the $\pi N$ S$_{11}$, S$_{31}$, and
P$_{11}$ partial-waves. Table~II compares the energy-dependent
and single-energy fits from threshold to 1.8 GeV.

In our most recent analysis\cite{aswp}
of elastic $\pi N$ scattering data, we
found evidence for two small structures on the high-energy tails
of the $S_{11}$(1650) and $F_{15}$(1680) resonances. These structures
remain small in the photoproduction reaction as well. In fact, they are
too small for a reliable estimate of their photo-decay amplitudes.

A set of $N \gamma$ decay couplings has been extracted from our
multipole amplitudes. We have fit these couplings using a
background plus Breit-Wigner form, as is described in  Ref.\cite{ar901}.
We analyzed both the energy-dependent and single-energy solutions
over a variety of energy ranges in order to estimate uncertainties.
Our results are listed in Table~III.
Here the resonance mass ($W_R$) and width ($\Gamma$) values
were obtained from fits to our multipole amplitudes. The values of $W_R$
remained quite consistent with estimates from our elastic $\pi N$
analysis. The results for $\Gamma$ tended to show more variation.
Values of $\Gamma_{\pi} /\Gamma$, where $\Gamma_{\pi}$ is the decay
width to $\pi N$ final states, were taken from the elastic $\pi N$
analysis and were not varied. This ratio is required in calculating the
photo-decay amplitudes.

As expected, there was little change in the photo-decay amplitudes for
resonances strongly coupled to $\pi N$ final states. These include the
$P_{33}$(1232), $D_{13}$(1520), $S_{11}$(1650),
$F_{15}$(1680), and $F_{37}$(1950). The $D_{15}$(1675) and $D_{33}$(1700),
have also remained stable. The most significant changes were found in
the $S_{11}$(1535) and $P_{11}$(1440) $A_{1/2}$ $\gamma n$ couplings.
As these resonances reflect complicated structures in the complex plane,
uncertainty in the $\gamma n$ coupling is not surprising. We should also
note that the $S_{11}$(1535) $\gamma p$ $A_{1/2}$ coupling remains
considerably below the value extracted from a recent analyses\cite{mukh} of
eta photoproduction data. A detailed analysis of both pion and eta
photoproduction data in this region would be useful.
A listing of our resonance couplings is given in Table ~III.

\section{SUM RULES}
\label{sec:srule}

The development of Chiral Perturbation theory (ChPT) and extended current
algebra has led to a renewed interest in a number of sum rules derived
in the 1960's. Examples include the Gerasimov-Drell-Hearn\cite{gdh} (GDH)
and Weinberg\cite{wsr} sum rules, as well as sum rules for the nucleon
electric, magnetic, and spin-dependent polarizabilities\cite{sand}.
Here we will briefly consider the status of two sum rules which require
input from photoproduction amplitudes. These are the GDH sum rule and
a sum rule\cite{ffr}, due to Fubini, Furlan and Rossetti (FFR),
which has not attracted as much attention.

While the
GDH sum rule was first derived from a dispersion relation (unsubtracted)
and the low-energy theorem (LET) for Compton scattering, it was later
obtained from the commutation relations of vector current densities.
In Ref.\cite{clw}, the extended current algebra of
Chang and Liang\cite{cl} was found to imply a modified GDH sum rule.
(It was observed\cite{ks} that
modified currents would lead to modified sum rules soon after the
original GDH sum rule appeared.)
An estimation of this modification was shown to account for the apparent
discrepancy\cite{wa} in the original sum rule.

In their discussion of modified sum rules, the authors of Ref.\cite{ks}
mentioned in passing that a similar procedure  could be used to determine
modifications to the FFR sum rule.
This sum rule relates nucleon magnetic moments to an integral over the
invariant amplitude ($A_1$) for single-pion photoproduction.
The FFR sum rule has the form\cite{diff}
\begin{equation}
g_A \left( { {e \kappa^{V,S}} \over {2 M} } \right) =
{ {2 f_{\pi} } \over {\pi} } \int Im \; A^{(+,0)}_1 (\nu) \; {{d\nu}\over \nu}
\end{equation}
where $\kappa^{V,S}$ is the isovector (isoscalar) anomalous
magnetic moment of the nucleon, given by $(\kappa_p \mp \kappa_n)/2$.
The invariant amplitude $A_1$ corresponds to the amplitude associated with
$\gamma_5 \gamma \cdot \epsilon \gamma \cdot k$
in the paper of Chew, Goldberger, Low, and Nambu\cite{cgln}.
The required isospin combinations are given\cite{cgln}, in terms of
charge-channel information, by
\begin{equation}
A_1^{(+,0)} = \left( A_1 \; (\gamma p\to \pi^0 p) \; \pm \;
                   A_1 \; (\gamma n\to \pi^0 n) \right) \; / \; 2.
\end{equation}
Here the amplitude for photoproduction of $\pi^0 n$ states is inferred from
measurements in the three other charge channels.

Empirical evaluation
of the integral in Eq.(1) is (in principle) much simpler than the
integral in the GDH sum rule $-$ which involves contributions from multi-pion
final states. Unfortunately, there are two problems which make a precise
check of the FFR sum rule more difficult.
Unlike the GDH sum rule, the FFR sum rule is not
exact.  It requires use of the Goldberger-Treiman relation\cite{gt}. In
addition, convergence of the associated integral is expected to be less
rapid than was found in the GDH sum rule.

Regardless of the above qualifications,
early attempts to evaluate the integral in Eq.(1) were encouraging. An
analysis\cite{ffr} using the $P_{33}(1232)$ and $D_{13}(1520)$ resonances
found good agreement for both $\kappa^V$ and $\kappa^S$.
A subsequent study\cite{ag},
using an early multipole analysis\cite{walk}, found 85\% of the prediction
for $\kappa^V$ but did not present results for the isoscalar combination.
In Ref.\cite{ag} the threshold behavior of the multipoles was modified
by a factor to account for a non-zero pion mass\cite{extrap}.

This brings us to the reason for re-examining the FFR sum rule.
If the FFR sum rule is valid, as the early studies suggest,
it puts a constraint on the contribution to the GDH sum rule
coming from single-pion photoproduction alone. Other tests of the
GDH integral (including the $\pi \pi N$ contributions)
have been made recently by Sandorfi et al.\cite{sand}.
In reference\cite{sand},
the multipole input to the GDH and spin-dependent polarizability
sum rules was compared
to predictions from ChPT\cite{meis}.
The integrals in these sum rules involve the difference of helicity 3/2
and 1/2 total cross sections
weighted by different powers of the photon energy.
The difference of
proton and neutron spin-dependent polarizabilities was found to agree with
ChPT while the difference of proton and neutron GDH sum rules is
known to have a problem\cite{wa}. The qualitative behavior found in
Refs.\cite{sand},\cite{wa} is preserved in the present analysis. The
isovector-isovector component  of the GDH sum rule receives a single-pion
production contribution very near the old estimate of Karliner\cite{wa}
while the isovector-isoscalar (VS) component retains its sign and magnitude
discrepancy.

While such comparisons are interesting, our poor knowledge of the $\pi \pi N$
contribution is an impediment. Early estimates of the $\pi \pi N$
contributions were based upon the resonance spectrum found in analyses of
$\pi N$ elastic scattering data. This neglects contributions from
possible "missing states" which couple very weakly to the $\pi N$ channel.
(Though the FFR sum rule is not exact, we at
least understand the approximation (PCAC) we are making.)

The integral giving $\kappa^V$ is heavily dominated by the
$P_{33} (1232)$ contribution, while the integral corresponding to
$\kappa^S$ appears to have important contributions from a wider range of
energies. The result for $\kappa^V$ was found to vary between
$1.8$ and $2.0$, remarkably close to
the predicted value. The integral corresponding to
$\kappa^S$, however, shows considerable sensitivity to uncertainties in
the high energy region. Here we find only qualitative agreement
(correct sign and order of magnitude). The energy dependence of the
isovector FFR integrand is displayed in Fig.~3.

In summary, we find the FFR sum rule for $\kappa^V$ to be well satisfied,
as was the case for isovector GDH sum rule. We also see that the FFR integral
does not converge as quickly as the analogous GDH integral.
The isoscalar result is less certain. The existence of significant
structure apart from the
$D_{13}$ resonance suggests that early success\cite{ffr}
with the isoscalar FFR component was fortuitous.
However, we should note that the isoscalar component of the FFR sum rule
appears to have less problems than the VS component of the GDH sum rule.
This tends to weaken arguments that require a large discrepancy in the
single-pion photoproduction multipoles in order to explain the
GDH discrepancy. It would be helpful if
high-quality photoproduction measurements could be extended
a further 1~GeV in order to
test the convergence of both the FFR and GDH sum rules.

If extended current algebra does indeed contribute
to the FFR sum rule (as suggested in Ref.\cite{ks}), the results
presented here should provide a useful test for the form
proposed by Chang and Liang\cite{cl}. While the isoscalar FFR sum rule
would likely provide the most sensitive check on any such contribution,
the phenomenological evaluation of the associated integral is not yet
sufficiently stable for more than an order-of-magnitude test.

\section{SUMMARY AND CONCLUSIONS}
\label{sec:sum}

We have extensively checked the pion photoproduction database for missing,
duplicated, and inconsistent measurements. This has resulted in a
significantly reduced $\chi^2$. The extracted photo-decay couplings generally
remain, for dominant resonances, in good agreement with the older
analysis of Crawford and Morton\cite{cm83}. The $\gamma n$ $A_{1/2}$
coupling for the $S_{11}$(1535) proved difficult to fit. The present
value is
quite different from the results of both Crawford and Morton\cite{cm83}
and our previous analysis\cite{larw} to 1.8~GeV.
The uncertainty in this coupling
is likely much greater than we previously estimated\cite{larw}. As mentioned
above, the $\gamma p$ coupling could also have a problem given the
discrepancy between the present value and the result of
eta photoproduction analyses.

The quark model results of Capstick\cite{cp92} reproduce most features of
the photo-decay couplings. The $P_{33}$(1232) couplings are underestimated,
but this is an old problem. The $P_{11}$(1440) couplings have the wrong
sign and magnitude. There have been suggestions\cite{lbl}
that this state, and also the $P_{33}$(1600), could be hybrids in which
case a comparison with the conventional quark model is inappropriate.
It is unfortunate that the weak resonance candidates, found in our analysis
of elastic pion-nucleon scattering data, were not clearly evident here.
These states should be considered in future analyses of other-meson
photoproduction databases.

We briefly examined two sum rules which require photoproduction input.
Those components dominated by the $P_{33}$(1232) resonance seem to be
reasonably well satisfied. The isoscalar components of the GDH sum rule
and the FFR sum rule for $\kappa^S$ are less certain. We are currently
exploring the use of fixed-t dispersion relations which may help to
constrain our analyses.

The results of these analyses, and the associated databases, are
available\cite{tel} via either Telnet or the Internet, or from the
authors upon request.

\acknowledgments

The authors express their gratitude to G.~Anton, B.~Bassalleck, R.~Beck,
M.~Blecher, H.~Dutz, K.~G.~Fissum, P.~Galumian, G.~Hakopian, M.~Khandaker,
M.~A.~Kovash, E.~Mazzucato, D.~F.~Measday, W.~Meyer, A.~S.~Omelaenko,
A.~M.~Sandorfi, P.~V.~Sorokin, J.~C.~Sta\v{s}ko, K.~Ukai, and
H.~B.~van~den~Brink for providing experimental data prior to publication or
for clarification of information already published.
R.W. acknowledges useful discussions regarding sum rules with Lay Nam Chang.
I.~S. acknowledges the hospitality extended by the Physics Department of
Virginia Tech.
This work was supported in part by the U.~S.~Department of Energy Grant
DE--FG05--88ER40454 and a NATO Collaborative Research Grant 921155U.


\eject

\newpage
{\Large\bf Figure captions}\\
\newcounter{fig}
\begin{list}{Figure \arabic{fig}.}
{\usecounter{fig}\setlength{\rightmargin}{\leftmargin}}
\item
{Energy-angle distribution of recent (post-1993) data.
(a) $\pi ^{0}p$, (b) $\pi ^{+}n$, and (c) $\pi ^{-}p$.
$\pi ^{0}p$ data are [observable (number of data)]:
d$\sigma$/d$\Omega$~(12), $\Sigma$~(111), T~(52), P~(6), O$_x$~(7), and
O$_z$~(7).
$\pi ^{+}n$ data are:
d$\sigma$/d$\Omega$~(261), $\sigma ^{tot}$~(3), $\Sigma$~(6), T~(222), and
P~(6).
$\pi ^{-}p$ data are:
d$\sigma$/d$\Omega$~(39), $\sigma ^{tot}$~(10), $\Sigma$~(9), T~(16), and
P~(10).
Total cross sections are plotted at zero degrees.}
\item
{Partial-wave amplitudes (L$_{2I, 2J}$) from 0 to 2~GeV.  Solid (dashed)
curves give the real (imaginary) parts of amplitudes corresponding to the
SM95 solution.  The real (imaginary) parts of single-energy solutions are
plotted as filled (open) circles.  The previous SP93 solution \cite{larw} is
plotted with long dash-dotted (real part) and short dash-dotted (imaginary
part) lines.  Plotted are the multipole amplitudes (a) $\rm _pE_{0+}^{1/2}$,
(b) $\rm _nE_{0+}^{1/2}$, (c) $\rm _pE_{0+}^{3/2}$, (d) $\rm _pM_{1-}^{1/2}$,
(e) $\rm _nM_{1-}^{1/2}$, (f) $\rm _pE_{1+}^{1/2}$, (g) $\rm _pM_{1+}^{1/2}$,
(h) $\rm _nE_{1+}^{1/2}$, (i) $\rm _nM_{1+}^{1/2}$, (j) $\rm _pM_{1-}^{3/2}$,
(k) $\rm _pE_{1+}^{3/2}$, (l) $\rm _pM_{1+}^{3/2}$, (m) $\rm _pE_{2-}^{1/2}$,
(n) $\rm _pM_{2-}^{1/2}$, (o) $\rm _nE_{2-}^{1/2}$, (p) $\rm _nM_{2-}^{1/2}$,
(q) $\rm _pE_{2-}^{3/2}$, (r) $\rm _pE_{2+}^{3/2}$, (s) $\rm _pE_{3-}^{1/2}$,
(t) $\rm _pM_{3-}^{1/2}$, (u) $\rm _nE_{3-}^{1/2}$, (v) $\rm _nM_{3-}^{1/2}$,
(w) $\rm _pE_{3-}^{3/2}$, and (x) $\rm _pM_{3+}^{3/2}$ in millifermi units.
The subscript p (n) denotes a proton (neutron) target.}
\item
{Integrand for the FFR sum rule giving $\kappa^V$.}
\end{list}

\begin{table}
\caption{Comparison of present (SM95)
and previous (SP93 and SP89)
energy-dependent partial-wave analyses of charged and neutral pion
photoproduction data.  $N_{prm}$ is the number parameters
varied in the fit.}
\label{tbl1}

\begin{tabular}{ccccccc}
Solution    &  Limit  & $\chi^2$/data & $\chi^2$/data
                      & $\chi^2$/data & $\chi^2$/data
& $N_{prm}$  \\
            &  (MeV)  & $\pi^0 p$  & $\pi^+ n$ &
     $\pi^- p$  &  $\pi^0 n$  &    \\
\tableline
SM95 & $2000$ & 13087/4711 & 12284/6359 & 6156/2225 & 282/120 & 135  \\
SP93~[1] & $1800$ & 14093/4015 & 22426/6019 & 8280/2312 & 275/120 & 134 \\
SP89~[7] & $1000$ & 13073/3241 & 11092/3847 & 4947/1728 & 461/120 & 97 \\
\end{tabular}
\end{table}
\newpage
\begin{table}
\caption{Comparison of single-energy (binned) and energy-dependent
analyses of pion photoproduction data.
$N_{prm}$ is the number parameters
varied in the single-energy fits. $\chi^2_E$ is due to the
energy-dependent fit (SM95) taken over the same energy interval.}
\label{tbl2}

\begin{tabular}{ccccccc}
E$_{lab}$~(MeV)&Range~(MeV)&$N_{prm}$&$\chi^2$/data&$\chi^2_E$&&\\
\tableline
 154 & $ 150 - 156 $ &  6 & 119/50  &  276 &&\\
 165 & $ 154 - 176 $ & 12 & 217/73  &  416 &&\\
 185 & $ 175 - 195 $ & 14 &  87/91  &  128 &&\\
 205 & $ 194 - 213 $ & 14 & 125/98  &  158 &&\\
 225 & $ 220 - 235 $ & 15 & 202/152 &  371 &&\\
 245 & $ 234 - 256 $ & 15 & 544/258 &  670 &&\\
 265 & $ 254 - 275 $ & 15 & 540/311 &  639 &&\\
 285 & $ 275 - 296 $ & 16 & 778/361 & 1000 &&\\
 305 & $ 294 - 316 $ & 16 & 722/431 &  866 &&\\
 325 & $ 314 - 336 $ & 17 & 902/423 & 1075 &&\\
 345 & $ 333 - 356 $ & 17 & 721/478 &  902 &&\\
 365 & $ 354 - 376 $ & 17 & 556/395 &  727 &&\\
 385 & $ 374 - 396 $ & 17 & 443/361 &  578 &&\\
 405 & $ 393 - 416 $ & 18 & 633/381 &  729 &&\\
 425 & $ 414 - 436 $ & 18 & 440/311 &  606 &&\\
 445 & $ 433 - 456 $ & 18 & 409/280 &  494 &&\\
 465 & $ 454 - 476 $ & 18 & 271/227 &  344 &&\\
 485 & $ 474 - 496 $ & 18 & 255/189 &  391 &&\\
 505 & $ 494 - 516 $ & 19 & 449/257 &  593 &&\\
 525 & $ 514 - 536 $ & 19 & 202/177 &  257 &&\\
 545 & $ 533 - 556 $ & 19 & 221/222 &  321 &&\\
 565 & $ 554 - 576 $ & 19 & 342/190 &  643 &&\\
 585 & $ 573 - 596 $ & 19 & 372/250 &  480 &&\\
 605 & $ 594 - 616 $ & 19 & 313/257 &  374 &&\\
 625 & $ 614 - 636 $ & 19 & 345/271 &  399 &&\\
 645 & $ 634 - 656 $ & 20 & 480/315 &  577 &&\\
 665 & $ 654 - 676 $ & 20 & 385/272 &  453 &&\\
 685 & $ 673 - 696 $ & 20 & 407/249 &  460 &&\\
 705 & $ 694 - 716 $ & 21 & 983/468 & 1139 &&\\
 725 & $ 714 - 736 $ & 21 & 290/221 &  468 &&\\
 745 & $ 733 - 756 $ & 21 & 766/409 & 1005 &&\\
 765 & $ 753 - 776 $ & 22 & 420/245 &  678 &&\\
 785 & $ 774 - 796 $ & 22 & 223/213 &  421 &&\\
 805 & $ 793 - 816 $ & 20 & 543/344 &  797 &&\\
 825 & $ 814 - 836 $ & 23 & 252/176 &  337 &&\\
 845 & $ 834 - 856 $ & 23 & 523/325 &  735 &&\\
 865 & $ 854 - 876 $ & 23 & 212/144 &  357 &&\\
 885 & $ 873 - 896 $ & 23 & 282/155 &  453 &&\\
 905 & $ 893 - 916 $ & 24 & 719/329 &  931 &&\\
 925 & $ 913 - 936 $ & 25 & 174/145 &  320 &&\\
 945 & $ 934 - 956 $ & 25 & 459/252 &  629 &&\\
 965 & $ 954 - 975 $ & 25 & 230/126 &  374 &&\\
 985 & $ 974 - 996 $ & 25 & 140/124 &  334 &&\\
1005 & $ 994 -1016 $ & 25 & 763/283 & 1051 &&\\
1025 & $1014 -1036 $ & 25 & 251/128 &  406 &&\\
1045 & $1034 -1056 $ & 25 & 394/195 &  622 &&\\
1065 & $1054 -1076 $ & 25 & 131/123 &  299 &&\\
1085 & $1074 -1096 $ & 25 &  92/97  &  286 &&\\
1105 & $1094 -1115 $ & 25 & 524/217 &  801 &&\\
1125 & $1115 -1136 $ & 25 & 140/98  &  283 &&\\
1145 & $1134 -1155 $ & 25 & 233/159 &  314 &&\\
1165 & $1154 -1176 $ & 25 & 127/97  &  199 &&\\
1185 & $1174 -1194 $ & 25 &  90/82  &  148 &&\\
1205 & $1194 -1216 $ & 25 & 276/174 &  433 &&\\
1225 & $1214 -1236 $ & 25 &  69/80  &  167 &&\\
1245 & $1234 -1255 $ & 25 & 168/104 &  249 &&\\
1265 & $1254 -1276 $ & 16 &  68/62  &  102 &&\\
1285 & $1275 -1296 $ & 16 &  31/40  &   84 &&\\
1305 & $1294 -1315 $ & 16 & 326/128 &  454 &&\\
1325 & $1314 -1335 $ & 16 &  52/45  &  129 &&\\
1345 & $1335 -1355 $ & 26 & 137/90  &  210 &&\\
1365 & $1355 -1375 $ & 16 &  37/36  &   92 &&\\
1385 & $1375 -1395 $ & 16 &  78/42  &  167 &&\\
1405 & $1395 -1416 $ & 26 & 496/136 &  669 &&\\
1425 & $1415 -1436 $ & 16 &  66/53  &  105 &&\\
1445 & $1435 -1456 $ & 26 & 104/78  &  148 &&\\
1465 & $1455 -1475 $ & 16 &  37/15  &   64 &&\\
1485 & $1474 -1495 $ & 16 &  63/32  &  121 &&\\
1505 & $1494 -1515 $ & 26 & 226/107 &  432 &&\\
1525 & $1515 -1535 $ & 16 &  69/33  &  148 &&\\
1545 & $1535 -1555 $ & 26 &  85/55  &  132 &&\\
1565 & $1555 -1575 $ & 16 &  18/17  &   39 &&\\
1585 & $1575 -1595 $ & 16 &  35/30  &   51 &&\\
1605 & $1595 -1616 $ & 26 & 122/92  &  217 &&\\
1625 & $1614 -1635 $ & 16 &  48/23  &   75 &&\\
1645 & $1635 -1655 $ & 16 & 199/79  &  243 &&\\
1665 & $1655 -1675 $ & 16 &  29/35  &   48 &&\\
1685 & $1675 -1695 $ & 16 &  20/28  &   37 &&\\
1705 & $1694 -1715 $ & 26 & 206/92  &  275 &&\\
1725 & $1715 -1735 $ & 16 &   9/14  &   18 &&\\
1745 & $1735 -1755 $ & 16 & 172/46  &  213 &&\\
1765 & $1754 -1775 $ & 16 &  49/34  &   65 &&\\
1785 & $1775 -1796 $ & 16 &  20/19  &   34 &&\\
1805 & $1795 -1815 $ & 16 & 224/75  &  308 &&\\

\end{tabular}
\end{table}

\newpage

Table III. Resonance couplings from a
Breit-Wigner fit to the SM95 solution [VPI],
the analysis of Crawford and Morton [CM83]\cite{cm83},
Arai and \hbox{Fujii} [AF82]\cite{af82},
recent quark model\cite{cp92} predictions [CAP92], and
an average from the Particle Data Group [PDG]\cite{pdg}.
A $\dagger$ indicates the quantity was not fitted.

\begin{table}
\label{tbl3}

\begin{tabular}{lccccc}
  &       & \multicolumn{2}{c}{\bf $\gamma p(GeV)^{-1/2} *10^{-3}$}
          & \multicolumn{2}{c}{\bf $\gamma n(GeV)^{-1/2} *10^{-3}$} \\
{\bf Resonance State} & {\bf Reference}  & $A_{1/2}$ & $A_{3/2}$
                                         & $A_{1/2}$ & $A_{3/2}$    \\
${\bf S_{11}(1535)} $  & VPI  & $60\pm 15$ &
                        & $-20\pm 35$ &                          \\
\hspace*{0.15in} $W_{R}=1525(10)~MeV$ & CM83
                        & $   65\pm 16 $ &
                        & $  -98\pm 26 $ &                          \\
\hspace*{0.15in} $\Gamma _{\pi}/\Gamma=0.31 $ & AF82
                        & $ 80\pm 7 $  &
                        & $ -75\pm 8 $ &                            \\
\hspace*{0.15in} $\Gamma = 103(5)~MeV $ & PDG
                        & $   68\pm 10 $ &
                        & $  -59\pm 22 $ &                          \\
                        & CAP92
                        & $   76 $ &
                        & $  -63 $ &                               \\
\hspace*{0.15in} &  &   &  &  &                                     \\
${\bf S_{11}(1650)} $  & VPI & $69\pm 5$ &
                        & $-15\pm 5$ &                         \\
\hspace*{0.15in} $W_{R}=1677(8)~MeV$ & CM83
                        & $   33\pm 15$ &
                        & $ -68\pm 40$ &                            \\
\hspace*{0.15in} $\Gamma _{\pi}/\Gamma \approx 1 $ & AF82
                        & $  61\pm 5  $ &
                        & $  8\pm 19  $ &                           \\
\hspace*{0.15in} $\Gamma =160(12)~MeV $ & PDG
                        & $   52\pm 17 $ &
                        & $  -11\pm 28 $ &                          \\
                        & CAP92
                        & $    54      $ &
                        & $   -35 $ &                               \\
\hspace*{0.15in} &  &   &  &  &                                     \\
${\bf P_{11}(1440)} $  & VPI & $-63\pm 5$ &
                        & $45\pm 15$ &                          \\
\hspace*{0.15in} $W_{R}=1463(7)~MeV$ & CM83
                        & $  -69\pm 18 $ &
                        & $   56\pm 15 $ &                          \\
\hspace*{0.15in} $\Gamma _{\pi}/\Gamma=0.68 $ & AF82
                        & $  -66\pm 4 $ &
                        & $   19\pm 12 $ &                          \\
\hspace*{0.15in} $\Gamma =360(20)~MeV $ & PDG
                        & $  -72\pm  9 $ &
                        & $   52\pm 25 $ &                          \\
                        & CAP92
                        & $    4      $ &
                        & $   -6   $ &                               \\
\hspace*{0.15in} &  &   &  &  &                                     \\
${\bf P_{11}(1710)} $  & VPI  & $7\pm 15$ &
                        & $-2\pm 15$ &                           \\
\hspace*{0.15in} $W_{R}=1720(10)~MeV$ & CM83
                        & $    6\pm 18 $ &
                        & $  -17\pm 20 $ &                          \\
\hspace*{0.15in} $\Gamma _{\pi}/\Gamma=0.15 $ & AF82
                        & $ -12\pm 5 $   &
                        & $ 11\pm 21 $   &                          \\
\hspace*{0.15in} $\Gamma =105(10)~MeV $ & PDG
                        & $   -6\pm 27 $ &
                        & $   16\pm 29 $ &                          \\
                        & CAP92
                        & $   13      $ &
                        & $   -11  $ &                               \\
\hspace*{0.15in} &  &   &  &  &                                     \\
${\bf P_{13}(1720)} $  & VPI  & $-15\pm 15$ & $7\pm 10$
                        & $7\pm 15$ & $-5\pm 25$           \\
\hspace*{0.15in} $W_{R}=1713(10)~MeV$ & CM83
                        & $   44\pm 66 $ & $  -24\pm 36 $
                        & $   -3\pm 34 $ & $   18\pm 28 $           \\
\hspace*{0.15in} $\Gamma _{\pi}/\Gamma=0.16 $ & AF82
                        & $  71\pm 10 $ &  $ -11 \pm 11 $
                        & $  1 \pm 38 $ &  $ -134 \pm 44 $          \\
\hspace*{0.15in} $\Gamma =153(15)~MeV $ & PDG
                        & $   27\pm 24 $ & $  -26\pm 10 $
                        & $   18\pm 29 $ & $  -33\pm 59 $           \\
                        & CAP92
                        & $   -11      $ & $  -31  $
                        & $    4      $ & $  11  $                 \\
\hspace*{0.15in} &  &   &  &  &                                     \\
${\bf D_{13}(1520)} $  & VPI  & $-20\pm 7$ & $167\pm 5$
                        & $-48\pm 8$ & $-140\pm 10$           \\
\hspace*{0.15in} $W_{R}=1516(10)~MeV$ & CM83
                        & $  -28\pm 14 $ & $  156\pm 22 $
                        & $  -56\pm 11 $ & $ -144\pm 15 $           \\
\hspace*{0.15in} $\Gamma _{\pi}/\Gamma=0.61 $ & AF82
                        & $ -32\pm 5 $   &  $ 162 \pm 3 $
                        & $ -71 \pm 11 $ & $  -148 \pm 9 $          \\
\hspace*{0.15in} $\Gamma =106(4)~MeV $ & PDG
                        & $  -22\pm 18 $ & $  163\pm 7 $
                        & $  -62\pm  6 $ & $ -137\pm 13 $           \\
                        & CAP92
                        & $   -15      $ & $ 134  $
                        & $   -38      $ & $-114  $                 \\
\hspace*{0.15in} &  &   &  &  &                                     \\
${\bf D_{15}(1675)} $  & VPI  & $15\pm 10$ & $10\pm 7$
                        & $-49\pm 10$ & $-51\pm 10$           \\
\hspace*{0.15in} $W_{R}=1673(5)~MeV$ & CM83
                        & $   21\pm 11 $ & $   15\pm 9  $
                        & $  -59\pm 15 $ & $  -59\pm 20 $           \\
\hspace*{0.15in} $\Gamma _{\pi}/\Gamma=0.38 $  & AF82
                        & $   6\pm 5  $  & $  29 \pm 4  $
                        & $   -25\pm 27 $ & $ -71\pm 26 $           \\
\hspace*{0.15in} $\Gamma =154(7)~MeV $ & PDG
                        & $   18\pm 10 $ & $   18\pm 9 $
                        & $  -50\pm 14 $ & $  -70\pm 6 $           \\
                        & CAP92
                        & $     2      $ & $   3  $
                        & $   -35      $ & $ -51  $                 \\
\hspace*{0.15in} &  &   &  &  &                                     \\
${\bf F_{15}(1680)} $  & VPI  & $-10\pm 4$ & $145\pm 5$
                        & $30\pm 5$ & $-40\pm 15$           \\
\hspace*{0.15in} $W_{R}=1679(5)~MeV$ & CM83
                        & $  -17\pm 18 $ & $  132\pm 10 $
                        & $   44\pm 12 $ & $  -33\pm 15 $           \\
\hspace*{0.15in} $\Gamma _{\pi}/\Gamma=0.68 $ & AF82
                        & $ -28\pm 3 $ & $ 115 \pm 12 $
                        & $ 26 \pm 5 $  & $ -24 \pm 9  $            \\
\hspace*{0.15in} $\Gamma =124(4)~MeV $ & PDG
                        & $  -14\pm 8 $ & $  135\pm 17 $
                        & $   27\pm 10 $ & $  -35\pm 11 $           \\
                        & CAP92
                        & $    -38      $ & $  56  $
                        & $    19      $ & $ -23  $                 \\
\hspace*{0.15in} &  &   &  &  &                                     \\
${\bf S_{31}(1620)} $  & VPI  & $35\pm 20$ & $            $
                        & $            $ & $            $           \\
\hspace*{0.15in} $W_{R}=1672(5)~MeV$ & CM83
                        & $   35\pm 10 $ & $            $
                        & $            $ & $            $           \\
\hspace*{0.15in} $\Gamma _{\pi}/\Gamma=0.29 $ &  AF82
                        & $ -26\pm 8 $ &
                        &              &                            \\
\hspace*{0.15in} $\Gamma =147(8)~MeV $ & PDG
                        & $   30\pm 14 $ & $            $
                        & $            $ & $            $           \\
                        & CAP92
                        & $   81      $ & $      $
                        & $            $ & $      $                 \\
\hspace*{0.15in} &  &   &  &  &                                     \\
${\bf P_{31}(1910)} $  & VPI  & $-2\pm 8$ & $            $
                        & $            $ & $            $           \\
\hspace*{0.15in} $W_{R}=1910^\dagger~MeV$ & CM83
                        & $   14\pm 30 $ & $            $
                        & $            $ & $            $           \\
\hspace*{0.15in} $\Gamma _{\pi}/\Gamma=0.26 $  & AF82
                        & $ -31\pm 4 $  &
                        &               &                           \\
\hspace*{0.15in} $\Gamma =250^\dagger~MeV $ & PDG
                        & $  13\pm 22 $ & $            $
                        & $            $ & $            $           \\
                        & CAP92
                        & $   -8      $ & $      $
                        & $            $ & $      $                 \\
\hspace*{0.15in} &  &   &  &  &                                     \\
${\bf P_{33}(1232)} $  & VPI  & $ -141\pm 5  $ & $-261\pm 5  $
                        & $            $ & $            $           \\
\hspace*{0.15in} $W_{R}=1232.5(0.5)~MeV$ & CM83
                        & $ -145\pm 15 $ & $ -263\pm 26 $
                        & $            $ & $            $           \\
\hspace*{0.15in} $\Gamma _{\pi}/\Gamma=0.99 $ & AF82
                        & $ -147 \pm 1 $ & $ -264 \pm 2 $
                        &                &                          \\
\hspace*{0.15in} $\Gamma =117(2)~MeV $ & PDG
                        & $ -141\pm  5 $ & $ -257\pm 8 $
                        & $            $ & $            $           \\
                        & CAP92
                        & $  -108      $ & $ -186  $
                        & $            $ & $       $                \\
\hspace*{0.15in} &  &   &  &  &                                     \\
${\bf P_{33}(1600)} $  & VPI  & $-18\pm 15 $ & $-25\pm 15  $
                        & $            $ & $            $           \\
\hspace*{0.15in} $W_{R}=1672(15)~MeV$ & CM83
                        & $  -39\pm 30 $ & $  -13\pm 14 $
                        & $            $ & $            $           \\
\hspace*{0.15in} $\Gamma _{\pi}/\Gamma=0.17 $ & AF82
                        &  -             & -
                        &                &                          \\
\hspace*{0.15in} $\Gamma =315(20)~MeV $ & PDG
                        & $  -26\pm 20 $ & $   -6\pm 17 $
                        & $            $ & $            $           \\
                        & CAP92
                        & $   30      $ & $  51  $
                        & $            $ & $       $                \\
\hspace*{0.15in} &  &   &  &  &                                     \\
${\bf D_{33}(1700)} $  & VPI  & $90\pm 25$ & $97\pm 20 $
                        & $            $ & $            $           \\
\hspace*{0.15in} $W_{R}=1690(15)~MeV$ & CM83
                        & $  111\pm 17 $ & $  107\pm 15 $
                        & $            $ & $            $           \\
\hspace*{0.15in} $\Gamma _{\pi}/\Gamma=0.16 $ & AF82
                        & $  112 \pm 6 $ & $ 47 \pm 7 $
                        &                &                          \\
\hspace*{0.15in} $\Gamma =285(20)~MeV $ & PDG
                        & $  114\pm 13 $ & $   91\pm 29 $
                        & $            $ & $            $           \\
                        & CAP92
                        & $    82      $ & $   68  $
                        & $            $ & $       $                \\
\hspace*{0.15in} &  &   &  &  &                                     \\
${\bf D_{35}(1930)} $  & VPI  & $-7\pm 10  $ & $ 5\pm 10  $
                        & $            $ & $            $           \\
\hspace*{0.15in} $W_{R}=1955(15)~MeV$ & CM83
                        & $  -38\pm 47 $ & $  -23\pm 80 $
                        & $            $ & $            $           \\
\hspace*{0.15in} $\Gamma _{\pi}/\Gamma=0.11 $ &  AF82
                        &  -            & -
                        &               &                           \\
\hspace*{0.15in} $\Gamma =350(20)~MeV $ & PDG
                        & $  -15\pm 17 $ & $  -10\pm 22 $
                        & $            $ & $            $          \\
                        & CAP92
                        & $     -      $ & $   -    $
                        & $            $ & $       $               \\
\hspace*{0.15in} &  &   &  &  &                                    \\
${\bf F_{35}(1905)} $  & VPI  & $22\pm 5 $ & $-45\pm 5  $
                        & $            $ & $            $          \\
\hspace*{0.15in} $W_{R}=1895(8)~MeV$ & CM83
                        & $   21\pm 10 $ & $  -56\pm 28 $
                        & $            $ & $            $          \\
\hspace*{0.15in} $\Gamma _{\pi}/\Gamma=0.12 $ & AF82
                        & $ 31\pm 9 $ &  $ -45\pm 6 $
                        &             &                            \\
\hspace*{0.15in} $\Gamma =354(10)~MeV $ & PDG
                        & $   37\pm 16 $ & $  -31\pm 30 $
                        & $            $ & $            $          \\
                        & CAP92
                        & $   26       $ & $  -1  $
                        & $            $ & $       $               \\
\hspace*{0.15in} &  &   &  &  &                                    \\
${\bf F_{37}(1950)} $  & VPI   & $-79\pm 6  $ & $-103\pm 6   $
                        & $            $ & $            $          \\
\hspace*{0.15in} $W_{R}=1947(9)~MeV$ & CM83
                        & $  -67\pm 14 $ & $  -82\pm 17 $
                        & $            $ & $            $          \\
\hspace*{0.15in} $\Gamma _{\pi}/\Gamma=0.49 $ & AF82
                        & $ -83 \pm 5 $ & $ -100 \pm 5 $
                        &                &                         \\
\hspace*{0.15in} $\Gamma =302(9)~MeV $ & PDG
                        & $  -85\pm 17 $ & $  -101\pm 14 $
                        & $            $ & $            $          \\
                        & CAP92
                        & $   -33      $ & $  -42  $
                        & $            $ & $       $               \\
\end{tabular}
\end{table}

\begin{references}
\bibitem{extrap} As the extrapolation $m_{\pi} \to 0$
is not well defined, we have used the physical photoproduction amplitudes.
\end{references}
\end{document}